\documentclass{webofc}
\usepackage[varg]{txfonts}   
\usepackage{float}

\newcommand{\Q}{{\cal Q}}
\newcommand{\bc}{B_c}
\newcommand{\X}{X\left( 3872 \right)}

\newcommand{\ie}{\textit{i.e.}}
\newcommand{\dd}{\mathrm{d}}

\newcommand{\Gmzero}{\Gamma_0^{\rm mol}}
\newcommand{\Gtzero}{\Gamma_0^{\rm tet}}
\newcommand{\pT}{p_T}
\newcommand{\raa}{R_{\rm AA}}

\newcommand{\psip}{\psi\left( 2S \right)}

\begin{document}
\title{Charmonium, $B_c$ and X(3872) Transport at the LHC}
%
%

\author{\firstname{Biaogang} \lastname{Wu}\inst{1}\fnsep\thanks{\email{bgwu@tamu.edu}} \and
   \firstname{Zhanduo} \lastname{Tang}\inst{1}\fnsep\thanks{\email{zhanduotang@tamu.edu}} \and
   \firstname{Min} \lastname{He}\inst{2}\fnsep\thanks{\email{minhephys@gmail.com}}\and
        \firstname{Ralf} \lastname{Rapp}\inst{1}\fnsep\thanks{\email{rapp@comp.tamu.edu}}
}

\institute{Cyclotron Institute and Department of Physics and Astronomy, Texas A$\&$M University, College Station, TX 77843-3366, USA
\and Department of Applied Physics, Nanjing University of Science and Technology, Nanjing~210094, China
          }

\abstract{%
   We deploy a kinetic-rate equation to evaluate the transport of $J/\psi$, $\psip$, $\bc$ and $\X$ in ultrarelativistic heavy-ion collisions 
  and compare their production yields to experimental data from the Large Hadron Collider. The rate equation has two main transport parameters, \ie, the equilibrium limit and reaction rate for each state.
   The temperature-dependent equilibrium limits include charm- and bottom-quark fugacities based on their initial production. 
   The reaction rates for charmonia, bottomonia and $\bc$ 
   rely on charm- and bottom-quark masses as well as binding energies from a thermodynamic $T$-matrix approach.
   For the $\X$ particle, its internal structure information is encoded in reaction rates and initial conditions in the hadronic phase via two different scenarios: a loosely bound hadronic molecule vs.~a compact diquark-antidiquark tetraquark. 
}
\maketitle
\section{Introduction}
\label{intro}

The production of charm and bottom hadrons in ultrarelativistic heavy-ion collisions (URHICs) has been intensely studied for several decades to help unravel the structure of the strongly-interacting matter at high temperature. Recent data on heavy quarkonia, \ie, charmonia and bottomonia, as well as $\bc$ and exotic particles have further fueled these efforts~\cite{ALICE:2019lga,CMS:2021znk,CMS:2022sxl}.
In addition, the internal structure of $\X$ particle -- a compact tetraquark with $cq$ anti/-diquark components vs.~a loosely bound hadronic molecule $D$ and $D^*$ meson -- has been conjectured to affect its yields in URHICs~\cite{Esposito:2016noz}. Here, we report comprehensive results for the nuclear modification factor for various quarkonia using our rate-equation approach~\cite{Grandchamp:2003uw,Zhao:2011cv,Du:2015wha}, as function of both centrality and transverse momentum 
($\pT$) and compare to pertinent data from the Large Hadron Collider (LHC). 
\section{Transport Approach in Thermal Medium}
\label{sec_trans} 
We focus on Pb-Pb collisions at 5\,TeV, simulating the medium evolution via a cylindrically expanding fireball with a transverse flow of blastwave type~\cite{Grandchamp:2003uw,Zhao:2011cv,Du:2015wha}. In this background, we solve rate equations including dissociation and regeneration for each quarkonium state, $\Q$, 
\begin{equation}
   \frac{dN_{\Q}(\tau)}{d\tau}=-\Gamma(T(\tau))\left[N_{\Q}(\tau)-N_{\Q}^{\rm eq}(T(\tau))\right] \ .
\label{eq_rateeq}
\end{equation}
This equation is governed by two transport parameters. The first one is the equilibrium limit, 
\begin{equation}
   N_{\Q}^{\rm eq}(T(\tau)) =  V_{\rm FB} d_{\Q} \gamma_{Q_1} \gamma_{Q_2} \int \frac{d^3k}{(2\pi)^3} \exp(-\sqrt{k^2+m_{\Q}^2}/T)     \  , 
\end{equation}
with the fireball volume, $V_{\rm FB}$, and the quarkonium degeneracy, $d_{\Q}$. The heavy-quark (HQ) fugacities, $\gamma_{Q_i}$ ($Q_i$=$b,c$), are calculated assuming heavy-flavor conservation within the thermal statistical model,
\begin{equation}
\label{Neq}
N_{Q\bar Q}=\frac{1}{2}\gamma_{Q} n_{\rm{op}}V_{\rm{FB}}\frac{I_1(\gamma_{Q} n_{\rm{op}}V_{\rm{FB}})}
{I_0(\gamma_{Q} n_{\rm {op}}V_{\rm{FB}})} + \gamma_{Q}^2 n_{\rm{hid}} V_{\rm{FB}}  \ , 
\end{equation}
where $N_{Q\bar Q}$ is the total number of charm-anticharm (or bottom-antibottom) pairs in the fireball, determined by hard production in primordial $NN$ collisions at given collision centrality, and $I_0$ and $I_1$ are the modified Bessel functions of zeroth and first order.

The inelastic reaction rates, $\Gamma_{\Q}$, for charmonia, bottomonia and $\bc$~are computed in quasi-free approximation~\cite{Grandchamp:2001pf}, with HQ masses, $m_{Q}$, 
and binding energies taken from in-medium $T$-matrix calculations~\cite{Liu:2017qah}. For the $\X$, we employ a schematic parameterization~\cite{Wu:2020zbx}
\begin{equation}
\Gamma(T)=\Gamma_0 \left( \frac{T}{T_0} \right)^n \ , 
\end{equation}
akin to hadronic reaction rates for charmonia~\cite{Du:2015wha} and focusing on $n$=3. With the guidance from existing literature~\cite{Cleven:2019cre,Brazzi:2011fq,Ferreiro:2018wbd}, $\Gamma_0$ at initial temperature of $T_0$=180\,MeV
are taken as $\Gmzero\simeq (400\pm100)$\,MeV and $\Gtzero\simeq 50$-80\,MeV for molecule and tetraquark, respectively, representing a hierarchy where the latter can survive in the QGP phase while the former cannot. This leads to an natural assumption for the initial conditions where the $\X$ abundance is at its chemical-equilibrium value for the tetraquark and zero for the molecule scenario~\cite{Wu:2020zbx}.

\section{Time Evolution and Observables}
\label{sec_obs}
\begin{figure}[!h]
	\begin{minipage}[b]{1.0\linewidth}
		\centering
                \includegraphics[width=0.46\textwidth]{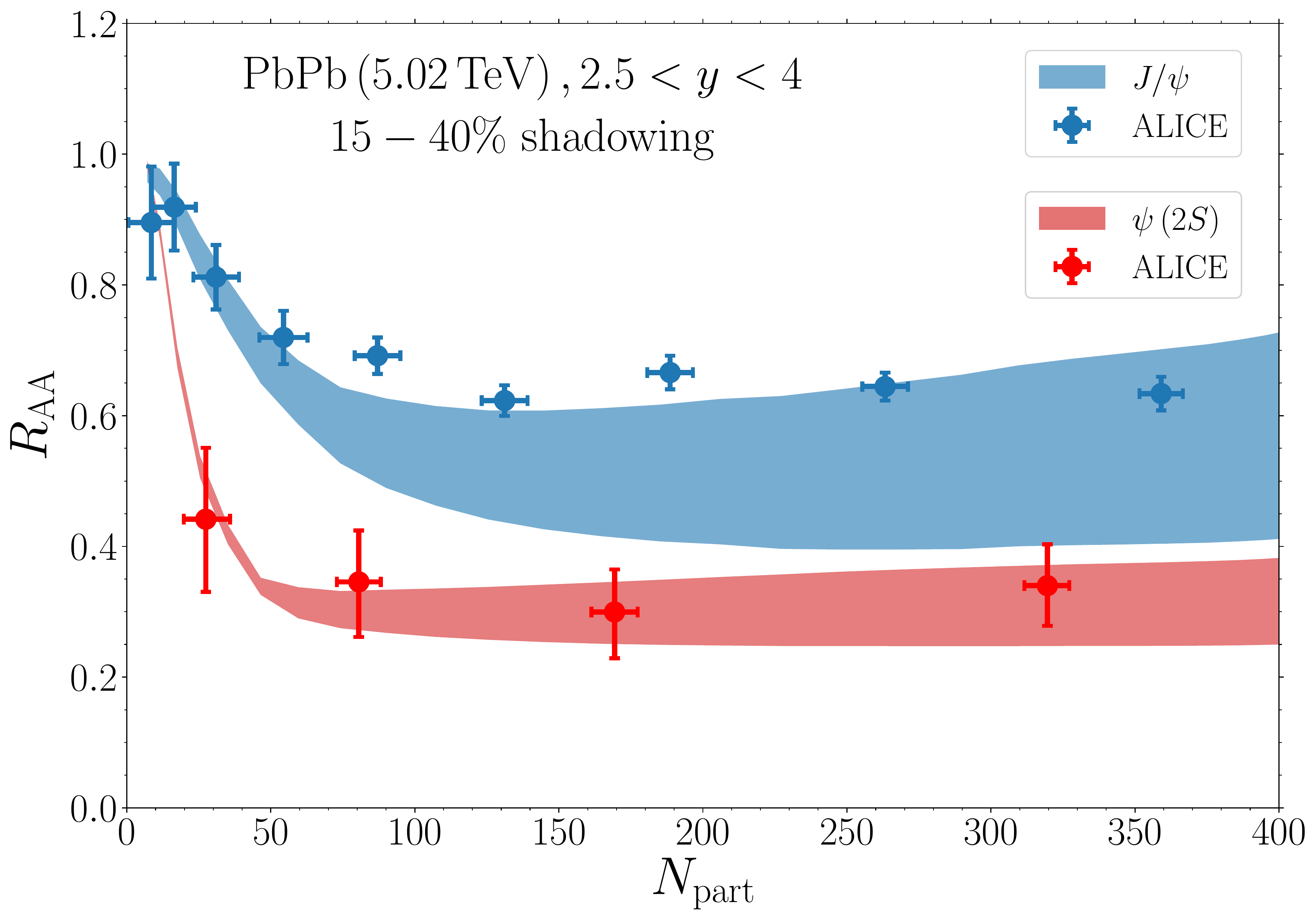}
                \includegraphics[width=0.46\textwidth]{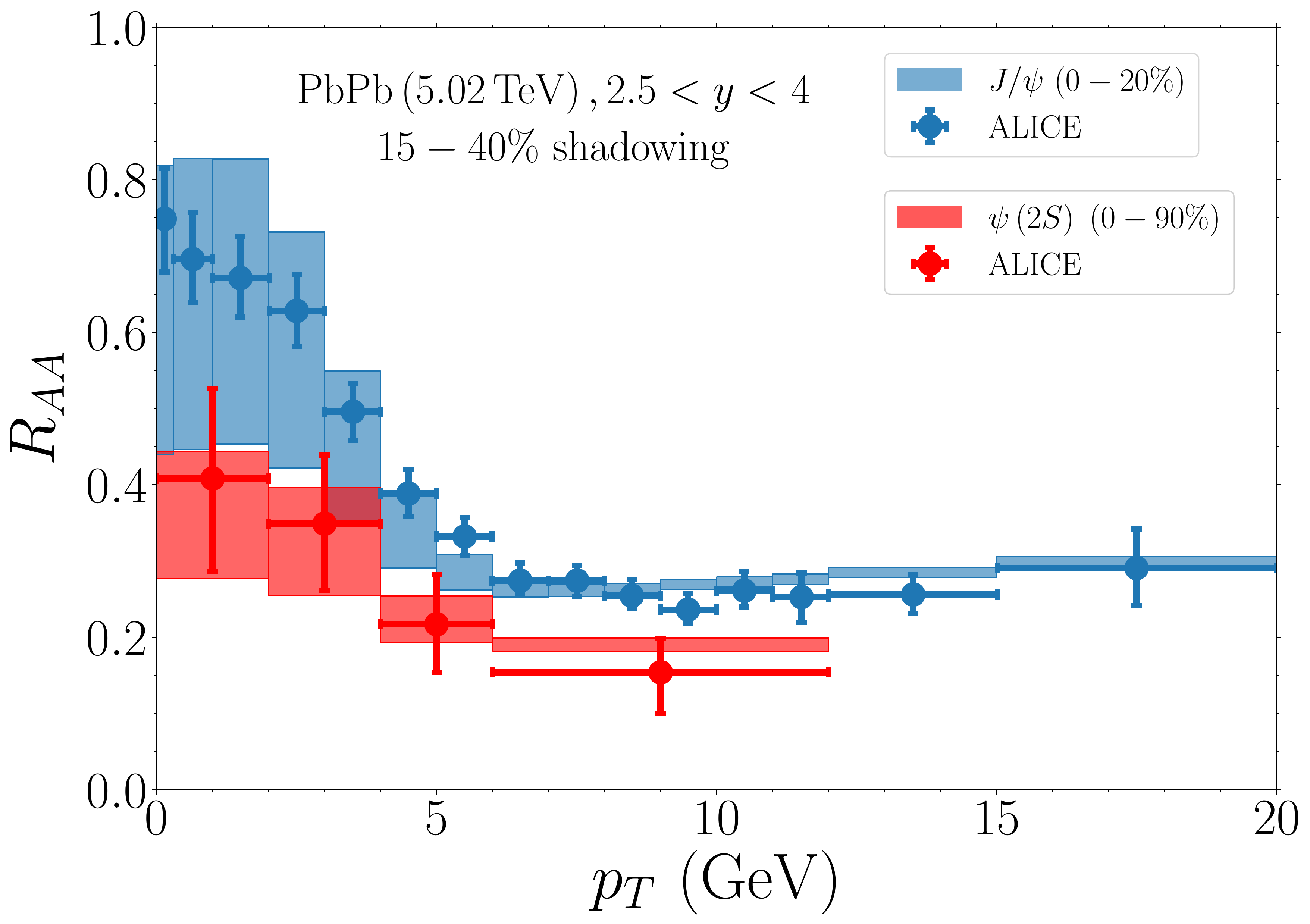}
	\end{minipage}
        \caption{Centrality (left) and $\pT$ (right) dependence for $J/\psi$ (blue) and $\psip$ (red bands) production in forward-rapidity 5.02\,TeV Pb-Pb collisions, compared to ALICE data~\cite{Hushnud:2022baz}.
        The bands represent the uncertainty due to the input $c\bar c$ cross section, $\dd\sigma_{c\bar{c}}^{pp}/dy$=0.72$\pm$0.7\,mb and 
        15-40\% shadowing.} 
	\label{fig_charm}
\end{figure}
Let us start with our calculations for charmonia using the most recent input charm/onium production cross sections from $pp$ collisions and shadowing estimates, cf.~Fig.~\ref{fig_charm}. The $\pT$-dependent $\raa$ for both charmonia show the characteristic signature of the maximum at low $\pT$. The centrality, and also the $\pT$ dependence show fair agreement with the most recent ALICE data, in particular given that the $\psip$ results were theoretical predictions~\cite{Hushnud:2022baz}.

\begin{figure}[!h]
\begin{minipage}[]{1.0\linewidth}
\centering
\includegraphics[width=0.46\textwidth]{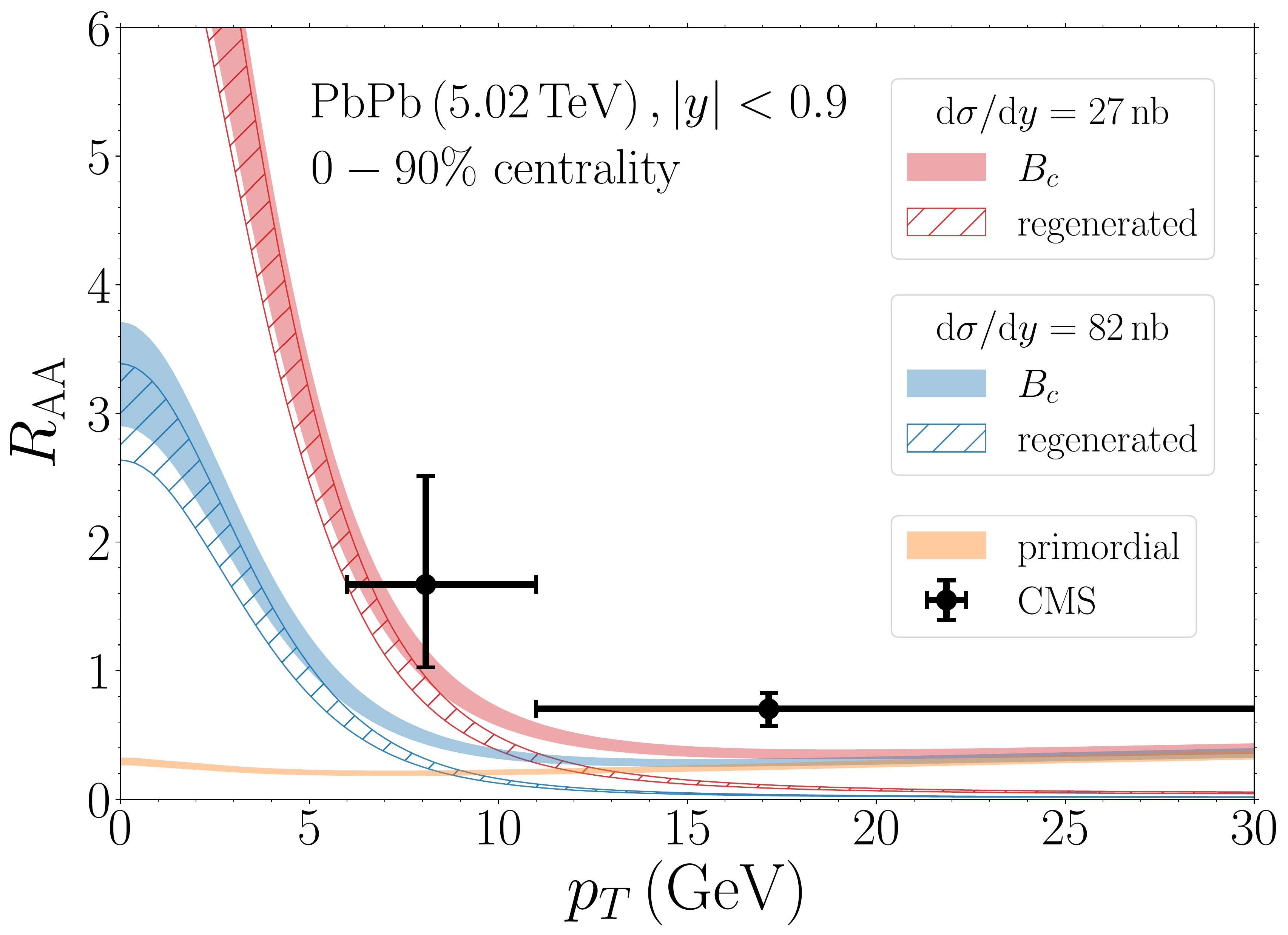}
\includegraphics[width=0.46\textwidth]{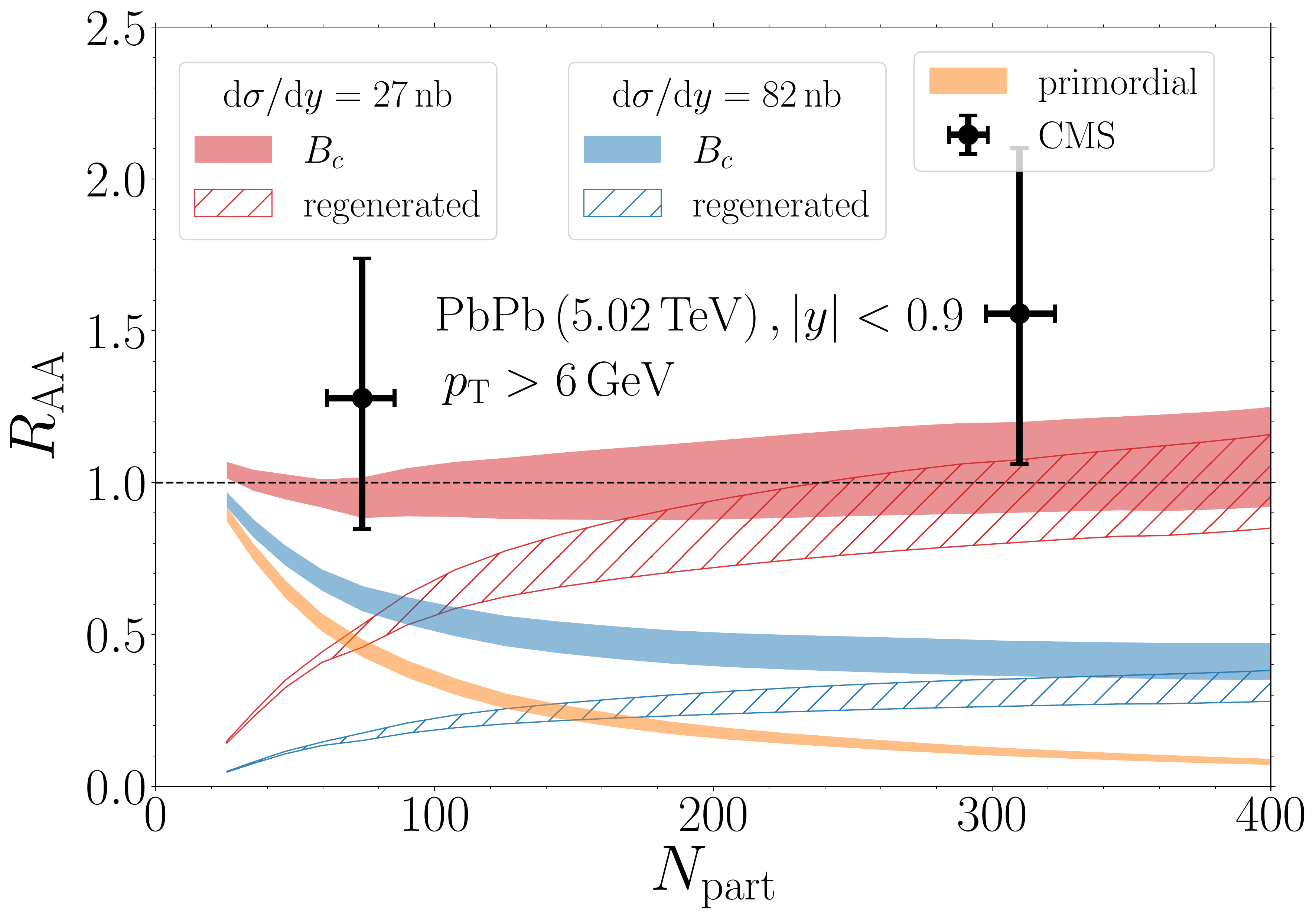}
\end{minipage}
\caption{Inclusive-$\bc$ $\pT$ spectra (left) and centrality dependence (right, with $\pT$$>$6\,GeV) in 0-90\% central Pb-Pb (5.02\,TeV) collisions, compared to CMS data~\cite{CMS:2022sxl}. The red (blue) curves correspond to a $pp$ cross section of $\dd\sigma/\dd y$=27 (82)\,nb; the primordial $\bc$ component (orange curves) is independent of the cross section in $pp$ collisions. The band widths reflect 10-30\% nuclear shadowing.}
\label{fig_bc}
\end{figure}
Next, we turn to our results for $\bc$ production, cf.~Fig.~\ref{fig_bc}. The input $pp$ cross section (figuring in the denominator of the $\raa$) is currently not well known; we have estimated it following Ref.~\cite{LHCb:2019tea} at $\dd\sigma^{pp}/\dd y$=27-82\,nb, with $\pT$ spectra extrapolated from 8\,TeV forward-rapidity data~\cite{LHCb:2014mvo}. 
We first compute the centrality dependence of the inclusive yields 
and then obtain the $\pT$ spectra for the regeneration contribution from recombining $b$- and $c$-quark spectra from Langevin transport calculations~\cite{He:2012xz}, at average formation temperatures for 
$\bc(1S)$ and $\bc(1P)$ of $T$=220\,MeV and $T_c$, respectively (the suppressed primordial spectra are obtained from Boltzmann simulations). The regenerated $\bc$ dominate the spectra up to 
$\pT\simeq 2m_{\bc}$. The comparison to CMS data~\cite{CMS:2022sxl},
especially for the centrality dependence with a $\pT>6$\,GeV cut, shows better agreement for smaller $pp$ cross sections.

\begin{figure}[H]
	\begin{minipage}[b]{1.0\linewidth}
		\centering
		\includegraphics[width=0.46\textwidth]{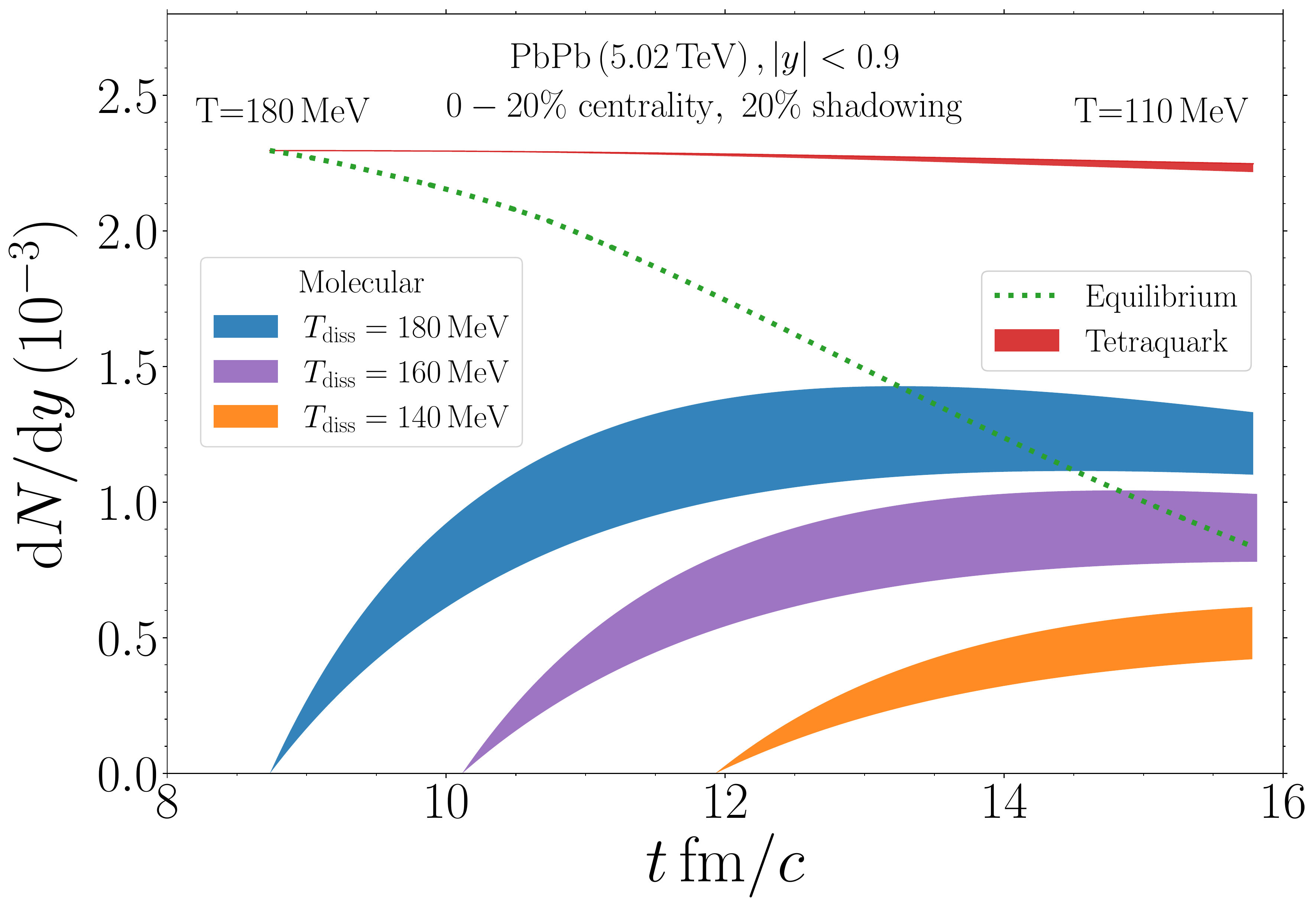}
		\includegraphics[width=0.46\textwidth]{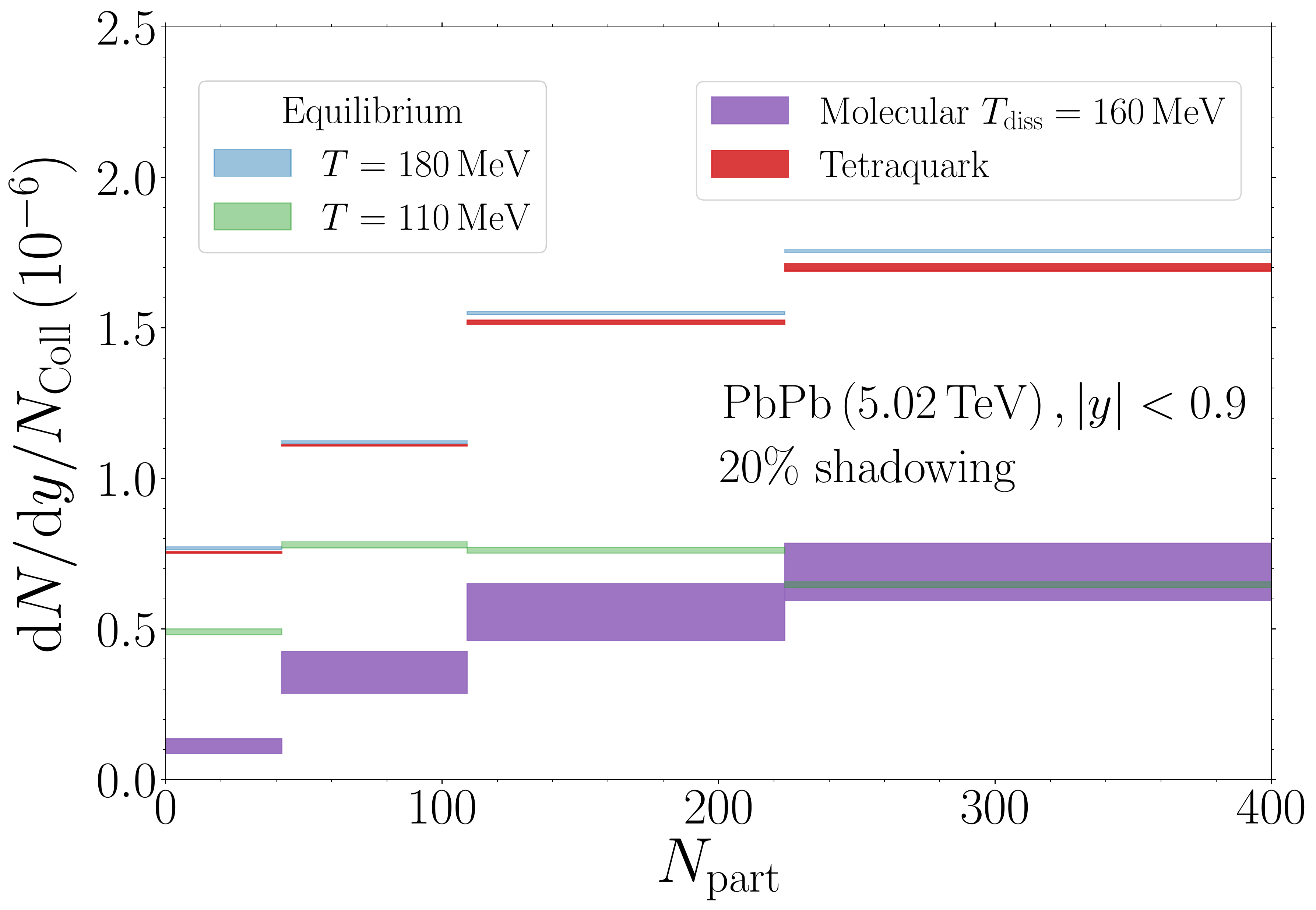}
	\end{minipage}
	\caption{$\X$ production in 0-20\% Pb-Pb (5.02\,TeV) collisions. Left: time evolution in hadronic matter for the molecule (lower 3 bands), tetraquark (red) and the equilibrium limit (green line). Right: centrality dependence for molecule (purple), tetraquark (red) and equilibriums limit at chemical (blue) and thermal freezeout (green); band widths reflect rates of  $\Gmzero$=300-500\,MeV and $\Gtzero$=50-80\,MeV.
} 
	\label{fig_X}
\end{figure}
Turning to $\X$ production, the time evolution of its yield in the hadronic phase (cf.~left panel of Fig.~\ref{fig_X}) shows little variation for the tetraquark scenario, remaining near its level at chemical freezeout. On the other hand, for the loosely bound hadronic molecule, assumed to come into existence only in the hadronic phase, appreciable regeneration occurs, quantitatively depending on its dissociation temperature. In any case, its final yield is significantly lower than for the tetraquark scenario.
The calculated centrality dependencies (normalized to the number of
binary nucleon-nucleon collisions), displayed in right panel of Fig.~\ref{fig_X}, show a rising trend for both scenarios -- a clear-cut signature of regeneration, which, however, is a factor of 2-5 larger for the (early produced) tetraquark compared to the (later produced) molecule.


\section{Conclusions}
\label{sec_concl}
We have conducted transport calculations of $J/\psi$, $\psip$, $\bc$~and $\X$ production in URHICs within a well-tested kinetic rate-equation approach.
Our predictions for the $\psip$ turn out to be in good agreement with recent ALICE data. For $\bc$ production we find regeneration to dominate while pertinent $\raa$'s show a strong sensitivity to the (currently uncertain) input cross section from $pp$ collisions.
For the $\X$, we encoded different structure scenarios in a scale hierarchy of the reaction rates, $\Gamma^{\rm mol} \gtrsim T_c \ge \Gamma^{\rm tet}$. The resulting yields for the tetraquark scenario are 2-5 times higher than for the molecule scenario, which differs from most coalescence model predictions.

\vspace{0.3cm}

\noindent 
{\bf Acknowledgments.}
This work is supported by the U.S.~NSF under grant nos.~PHY-1913286 and PHY-2209335.

%
\bibliography{refcnew}
%
%
%
%

\end{document}